# Heteroepitaxial Growth and Multiferroic Properties of Mn-doped BiFeO$_3$ films on SrTiO$_3$ buffered III-V Semiconductor GaAs

G. Y. Gao[1,2], Z. B. Yang[1], W. Huang[1,3], H. Z. Zeng[3], Y. Wang[1], H. L. W. Chan[1], W. B. Wu[2], and J. H. Hao[1,a)]

[1]*Department of Applied Physics, The Hong Kong Polytechnic University, Hong Kong, P.R. China*

[2]*Hefei National Laboratory for Physical Sciences at Microscale, University of Science and Technology of China, Hefei, Anhui 230026, P. R. China*

[3]*State Key Laboratory of Electronic Thin Films and Integrated Devices, University of Electronics Science and Technology of China, Chengdu, P. R. China*

Abstract

Epitaxial Mn-doped BiFeO$_3$ (MBFO) thin films were grown on GaAs (001) substrate with SrTiO$_3$ (STO) buffer layer by pulsed laser deposition. X-ray diffraction results demonstrate that the films show pure (00$l$) orientation, and MBFO (100)//STO(100), whereas STO (100)//GaAs (110). Piezoresponse force microscopy images and polarization versus electric field loops indicate that the MBFO films grown on GaAs have an effective ferroelectric switching. The MBFO films exhibit good ferroelectric behavior ($2P_r \sim 92$ μC/cm$^2$ and $2E_C \sim 372$ kV/cm). Ferromagnetic property with saturated magnetization of 6.5 emu/cm$^3$ and coercive field of about 123 Oe is also found in the heterostructure at room temperature.

---------------------------------------------------------

[a)] Authors to whom correspondence should be addressed. Electronic email: jh.hao@polyu.edu.hk

**I. Introduction**

There has been a long-standing research interest in the integration of functional oxides with semiconductors not only for the fundamental physics, but also due to their potential applications in future technological devices. Generally, many related studies have focused on the use of silicon wafer for the integration of functional materials[1,2] and devices,[3-7] including ferroelectric field effect transistor,[3,4] multiferroic-based spin valve.[5] These systems integrated on silicon wafer have demonstrated a great achievement in this field. Recently, compound semiconductors, such as gallium nitride (GaN), are used as substrates for the growth of ferroelectric perovskite oxides driven by the potential application in multifunctional devices due to their superior physical properties.[8] For example, heteroepitaxy of BFO on GaN has been achieved.[4,5,9] However, there has been less attention attracted on the integration with III-V compound GaAs, which is one of the most important semiconductor materials besides silicon. Compared to Si, the structure and chemistry on the GaAs surface are more complex and involves multiple reconstructions, depending upon the surface termination and process conditions. Unfortunately, the integration of functional oxides into the III-V's semiconductor technology still represents a significant challenge. A few reports have ever attempted to integrate functional oxides with GaAs for exploiting novel electronic device. Ferroelectric oxide $BaTiO_3$ grown on GaAs (001) have been achieved by molecular beam epitaxy using MgO as a template,[10,11] but the properties are greatly dependent on interface structure. Recently, we have epitaxially grown SrTiO$_3$ (STO) on GaAs (001), and therefore get a more stable and high quality interface for further integration of functional oxdes.[12-14]

Bismuth ferrite BiFeO$_3$ (BFO) has been studied extensively due to its interesting ferroelectric polarization, high Curie temperature, especially the prominent multiferroism. Recently, some results indicate that BFO can also exhibit photovoltaic effect,[15,16] indicating the potential application of BFO in optoelectronic devices integrated with GaAs, which exhibits a direct energy band gap and superior optical properties. Bulk BFO has a pseudo-cubic perovskite structure with a lattice constant of 3.96 Å,[17] which can be grown on some semiconductors by selecting suitable buffer layers. However, there has been no report on the investigation of BFO films grown on GaAs semiconductor. In this work, we report on the epitaxial growth of multiferroic films of Mn-doped (5%) BiFeO$_3$ (MBFO) on GaAs (001) with STO as a template and multiferroic properties of the heterostructure are observed at room temperature.

**II. Experimental**

MBFO films were grown on GaAs (001) substrate by pulsed laser deposition (PLD) using an excimer laser ($\lambda$=248 nm) via a 50 nm thick STO (001) thin film as buffer layer. The high quality STO film can be grown on GaAs (001) directly at 590 $^o$C, the experimental details can be found in our previous reports.[12-14] After the buffer layer deposition, the substrate temperature was fixed at 620-650 $^o$C, and the oxygen pressure was raised up to 9 Pa for subsequent MBFO deposition. During deposition, the laser frequency and energy density were kept at 10 Hz and 6 J/cm$^2$, respectively.

The distance between the target and substrate was 5 cm. The thickness of the MBFO was about 250 nm. After MBFO deposition, the sample was *in-situ* annealed at 420 °C in 100 Pa partial oxygen pressure before cooled down to room temperature in order to compensate oxygen vacancies in the films.[18] Finally, Au electrode with a diameter of 0.2 mm was sputtered onto the MBFO films using a shadow mask. The crystal structure of the samples was examined by X-ray diffraction technique. The surface morphology and piezoelectric force microscopy (PFM) measurements were carried out on a scanning probe microscope (SPM). Ferroelectric polarization was characterized using a ferroelectric tester. The measurement of frequency dependent dielectric constant and loss was performed at room temperature using an Agilent 4284A Precision LCR Meter. Magnetic properties were measured at room temperature using a vibrating sample magnetometer (Quantum Design, SQUID VSM).

III. Results and Discussion

The samples grown in temperature range of 620-650 °C exhibit a single phase and have an epitaxial relationship with GaAs substrate. Figure 1 (a) shows a typical XRD linear scan on a semi-logarithmic scale for MBFO films grown on STO buffered GaAs (001) substrate along the normal of GaAs (001) specular reflections. Only (00*l*) reflections from MBFO films were observed besides for the GaAs (00*l*) reflections. No second phase appeared in the scans, indicating that the films have a single (00*l*) orientation. The lattice constant of the films was calculated to be 3.959 Å, comparable to that of bulk BFO, indicating fully relaxed strain state in the films. No STO peaks were observed in the scan because of the overlap with the peaks of MBFO films and lower intensity of the peaks due to thinner thickness of STO. To prove epitaxial nature of the MBFO films grown on GaAs (001) substrate, polar phi-scan (Fig. 1(b)) was used to examine the epitaxial relationship between MBFO and GaAs (001) substrate. The phi-scan peak alignment indicates the in-plane orientation relationship as MBFO (100) // STO (100) // GaAs(110), which is consistent with previous results.[12,13] The epitaxial relationship between BFO, STO, and GaAs unit cells is schematic shown in Figure 1 (c). The full-width at half-maximum (FWHM) of the rocking curve for the MBFO films was estimated at 1.2º (not shown), indicating limited crystallinity of BFO film grown at low temperature. We note that the FWHM value of our sample is still larger compared to other reports about BFO grown on other substrates such as single-crystal oxides.[19,20] However, when further increasing the substrate temperature, the improvement in crystallinity of MBFO films will be hampered by the sublimation of As atoms from GaAs, leading to a second orientation in MBFO films. One possibility leading to the large FWHM of our sample is the lattice deformation induced by the large lattice mismatch between GaAs and BFO or STO, although the buffered STO films were grown on GaAs with 45º rotated unit cell. On the other hand, the difference in the FWHM is likely to result from different deposition temperature and rate during BFO film deposition.

To understand the surface and interface structure of the sample, SPM was used to characterize the 1×1 μm surface of MBFO films. Figure 2(a) shows the topography of the as-grown MBFO film on GaAs (001) substrate. One can see that MBFO surface is distributed with 3D islands randomly nucleated on the surface,[21] and the root mean square roughness of the surface is about 7.33 nm. The ferroelectric properties of the sample were examined by PFM. To measure electrical properties of the MBFO film, a conductive Nb-doped (10%) STO (NSTO) film was used to replace STO deposited on GaAs as a bottom electrode. The amplitude and phase signals of vertical PFM were simultaneously obtained on the surface of the samples at same area, as shown in Figure 2 (b) and (c), respectively. From these images, a color contrast corresponding to a poly-domain state could be clearly seen under the PFM measurement with an ac bias voltage of 1v. The strong dark and bright contrast of the vertical PFM (VPFM) image (Fig2.(b)) implies a large polarization perpendicular to the BFO film. Piezoelectric contrast can also be switched with a DC voltage applied to the tip. The out-of-plane polarization was reversed effectively when bias voltage switched from 35 V to -35 V (Fig. 2(d)). Besides, a slight offset of the PFM hysteresis loop is observed, which could be attributed to the clamping effect from substrate.[22,23] The PFM hysteresis loop from out-of-plane was investigated as a function of the applied electric field. In the measurement, the ac voltage was 1 V at 46 kHz. The measured coercive electric-voltage ($V_c$) is about 13 V, indicating the ferroelectric behavior of spontaneous polarization in MBFO films grown on GaAs at nanoscale.

To further elucidate the ferroelectric properties of the MBFO films grown on GaAs (001), ferroelectric hysteresis loops (*P-E*) and the typical leakage current as function of applied voltage (*I-V*) were measured from the Au/MBFO/NSTO/GaAs capacitors as shown in Fig. 3. Two dynamic *P-E* loops of MBFO film deposited at 620 °C were recorded when the maximum driving voltages were 15 V and 25 V at 1 kHz using a bipolar pulse, as shown in Figure 3 (a). The polarization is lower than that from other reported works,[24,25] which is caused by low crystallinity in MBFO because of the use of lower growth temperature compared to the earlier reports. Moreover, the *P-E* loops do not show saturation which are often dominated by leakage currents caused by oxygen vacancies or the mixed valence nature of the Fe. Such behaviours imply that it is not preferable to switch inhomogeneous domains with different coercivity in the film under applied field.[26,27] For those films grown on STO or other oxide substrates, the crystallinity of the BFO films is generally to be improved by an appropriate increase of deposition temperatures or annealing temperatures.[28] For the films on GaAs, however, the use of excessively high growth temperature may lead to chemical interdiffusion at interface of STO/GaAs, and as temperature over 690 °C, the sublimation is also occurred from GaAs substrate.[12,13] All these processes can result in the degradation of physical properties of the heterostructure. Keeping in mind that the deposition temperature must be not higher than around 690 °C, we adopted a higher deposition temperature about 650 °C in order to optimize the MBFO growth condition. Figure 3 (b) shows the *P-E* loops from BFO/NSTO/GaAs heterojunction deposited at 650 °C. The *P-E* loops have been significantly improved, exhibiting better ferroelectric behavior ($2P_r \sim 92$ μC/cm² and $2E_C \sim 372$ kV/cm) at the applied electric field of 600 kV/cm. Moreover, a relative more saturation polarization is observed compared to the result in Figure 3 (a).

Additionally, the capacitors exhibit a large voltage offset in the *P-E* hysteresis loops, which is caused by some effects, including asymmetrical electrodes used in the heterostructure as well as an internal electric field in the capacitor producing an anti-parallel polarization state.[29] The measurements of frequency dependence of dielectric constant and loss tangent from MBFO film on NSTO buffered GaAs (001) substrate were measured with parallel-plate electrode structure at room temperature, as shown in Figure 3 (c) and (d). Over the whole range of frequency, both dielectric constant and loss tangent decrease as frequency increasing, suggesting that the dipoles are capable of following the frequency of the applied field at low frequencies.[30,31] The dielectric constant of the MBFO films is comparable to the value of other doped BFO films.

To examine the conduction mechanism in those *P-E* loops, the leakage current curves of the BFO/NSTO/GaAs capacitors were measured, as shown in Figure 3 (e). The applied electric field is swept at a fixed rate from zero bias to 300 kV/cm, then to −300 kV/cm before returning to zero. A slight difference was observed in leakage currents when the bias was reversed. The leakage current density $J$ is low, not exceed 1 mA/cm$^2$ at an applied electric field about 150 kV/cm, and the *J-E* curve of negative and positive biases also exhibits obvious asymmetry. To gain an insight into the nature of the leakage mechanism,[32] we have examined the leakage data in various manners as a function of applied electric field according to different leakage mechanisms. Note that log($J$) vs log($E$) at a negative bias for the heterostructure are linear with a slope about 1 at lower bias, indicating a normal ohmic conductive behavior. However, at higher electric field range, a linear relation between log($J/E$) vs $E^{1/2}$ is observed, as shown in Figure 3(f), suggesting the leakage current is governed by Poole-Frenkel (PF) emission mechanism with consecutive hopping of charges between defect trap centers.[33] These results demonstrate that the leakage current in *P-E* loops of BFO/NSTO/GaAs is dominated by PF mechanism caused by the defects of the system which could be further investigated by high-resolution transmission electron microscopy.[34]

Figure 4 shows the magnetic-field-dependent magnetization of the MBFO/NSTO/GaAs film measured at room temperature (300 K) when magnetic field is applied parallel to the sample surface. The magnetic signals from GaAs substrate and NSTO films were subtracted already from the raw data. The loop shows clearly hysteresis effect, indicating the weak ferromagnetic properties of the MBFO films. The in-plane saturated magnetization and the coercive field is about 6.5 emu/cm$^3$ and 123 Oe, which is comparable to the value reported by other works.[35]

## IV. Conclusion

In conclusion, we have deposited epitaxial Mn-doped BFO films on STO buffered GaAs (001) by PLD. X-ray diffraction results revealed that the heterostructure has in-plane relationship of MBFO (100) // STO (100)//GaAs (110). Ferroelectric switching is clearly observed by PFM. The polarization data provide an evidence of ferroelectricity with $2P_r$ ~ 92 μC/cm$^2$ and $2E_C$ ~ 372 kV/cm. Magnetic measurement results show that the films are ferromagnetic with saturated magnetization of 6.5 emu/cm$^3$ and coercive field of about 123 Oe. It is hoped that the multiferroic /III-V semiconductor heterostructure fabricated in this work will become an important component in multifunctional devices for many applications.

**Acknowledgements**

This work was supported by the grants from the Research Grants Council of Hong Kong (GRF Project No. PolyU500910) and ITS 029/11. Also, partial support by the Natural Science Foundation of China Grant No. 51002022 and No. 51002023 is acknowledged.

**Figure Captions**

Figure 1 (a) XRD linear scan for the MBFO films grown on STO buffered GaAs (001) substrate. (b) Phi-scan on the MBFO (101) and GaAs (202) reflections, respectively. (c) Schematic illustration of the epitaxial relationship between BFO, STO, and GaAs unit cells.

Figure 2 (a) Surface morphology of the MBFO/NSTO/GaAs heterostructure; (b) amplitude and (c) phase PFM images of the BFO/NSTO/GaAs heterostructure; (d) the piezoresponse output of the BFO/NSTO/GaAs heterostructure as a function of the applied electric voltage.

Figure 3 Room temperature ferroelectric hysteresis loops recorded at 1 kHz from the MBFO films deposited at 620 $^{o}$C (a) and 650 $^{o}$C (b) on GaAs (001), respectively. The loops is formed with 10 V at first, then with 25 V. The frequency dependent (c) dielectic constant and (d) dielectric loss of the MBFO films on NSTO/GaAs at room temperature. (e) Typical *J-E* curves of the MBFO/NSTO/GaAs capacitor for both negative and positive biases. (f) log(*J/E*) vs $E^{1/2}$ at negative bias. The solid line is fitting use Poole-Frenkel (PF) emission mechanism.

Figure 4 Magnetic-field-dependent magnetization of the MBFO on NSTO buffered GaAs measured at room temperature when magnetic field was applied parallel to the sample surface.

Reference

1. R. A. McKee, F. J. Walker, and M. F. Chisholm, Phys. Rev. Lett. **81**, 3014 (1998).

2. J. H. Hao, J. Gao, Z. Wang, and D. P. Yu, Appl. Phys. Lett. **87**, 131908 (2005).

3. T. Zhao, S. B. Ogale, S. R. Shinde, R. Ramesh, R. Droopad, J. Yu, K. Eisenbeiser, and J. Misewich, Appl. Phys. Lett. **84**, 750 (2004).

4. S. Y. Yang, Q. Zhan, P. L. Yang, M. P. Cruz, Y. H. Chu, R. Ramesh, Y. R. Wu, J. Singh, W. Tian, and D. G. Schlom, Appl. Phys. Lett. **91**, 022909 (2007).

5. L. W. Martin, Y.-H. Chu, Q. Zhan, R. Ramesh, S.-J. Han, S. X. Wang, M. Warusawithana, and D. G. Schlom, Appl. Phys. Lett. **91**, 172513 (2007).

6. C. W. Bark, D. A. Felker, Y. Wang, Y. Zhang, H. W. Jang, C. M. Folkman, J. W. Park, S. H. Baek, H. Zhou, D. D. Fong, X. Q. Pan, E. Y. Tsymbal, M. S. Rzchowski, and C. B. Eom, PNAS **108**, 4720 (2011).

7. S. H. Baek, J. Park, D. M. Kim, V. A. Aksyuk, R. R. Das, S. D. Bu, D. A. Felker, J. Lettieri, V. Vaithyanathan, S. S. N. Bharadwaja, N. Bassiri-Gharb, Y. B. Chen, H. P. Sun, C. M. Folkman, H. W. Jang, D. J. Kreft, S. K. Streiffer, R. Ramesh, X. Q. Pan, S. Trolier-McKinstry, D. G. Schlom, M. S. Rzchowski, R. H. Blick, and C. B. Eom, Science **334**, 958 (2011).

8. R. Chau, S. Datta, M. Doczy, B. Doyle, B. Jin, J. Kavalieros, A. Majumdar, M. Metz, and M. Radosavljevic, IEEE Trans. Nanotechnol. **4**, 153 (2005).

9. W. B. Luo, J. Zhu, H. Z. Zeng, X. W. Liao, H. Chen, W. L. Zhang, and Y. R. Li, J. Appl. Phys. **109**, 104108 (2011).

10. T. E. Murphy, D. Chen, and J. D. Phillips, Appl. Phys. Lett. 85, 3208 (2004).

11. D. Chen, T. E. Murphy, S. Chakrabarti, and J. D. Phillips, Appl. Phys. Lett. **85**, 5206 (2004).

12. Z. P. Wu, W. Huang, K. H. Wong, and J. H. Hao, J. Appl. Phys. **104**, 054103 (2008).

13. W. Huang, Z. P. Wu, and J. H. Hao, Appl. Phys. Lett. **94**, 032905 (2009).

14. W. Huang, J. Y. Dai, and J. H. Hao, Appl. Phys. Lett. **97**, 162905 (2010).

15. S. Y. Yang, L. W. Martin, S. J. Byrnes, T. E. Conry, S. R. Basu, D. Paran, L. Reichertz, J. Ihlefeld, C. Adamo, A. Melville, Y.-H. Chu, C.-H. Yang, J. L. Musfeldt, D. G. Schlom, J. W. Ager, III, and R. Ramesh, Appl. Phys. Lett. **95**, 062909 (2009).

16. T. Choi, S. Lee, Y. J. Choi, V. Kiryukhin, and S.-W. Cheong, Science **324**, 63 (2009).

17. Y. H. Chu, L. W. Martin, Q. Zhan, P. L. Yang, M. P. Cruz, K. Lee, M. Barry, Ferroelectrics **354**, 167 (2007).

18. H. Yang, M. Jain, N. A. Suvorova, H. Zhou, H. M. Luo, D. M. Feldmann, P. C. Dowden, R. F. DePaula, S. R. Foltyn, and Q. X. Jia, Appl. Phys. Lett. **91**, 72911 (2007).

19. D. Y. Wang, N. Y. Chan, R. K. Zheng, C. Kong, D. M. Lin, J. Y. Dai, H. L. W. Chan, and S. Li, J. Appl. Phys. **109**, 114105 (2011).

20. Y. Wang and J. Wang, J. Appl. Phys. **106**, 94106 (2009).

21. J. L. Li, J. H. Hao, Z. Ying, and Y. R. Li, Appl. Phys. Lett. **91**, 201919 (2007).

22. W. G. Chen, W. Ren, L. You, Y. R. Yang, Z. H. Chen, Y. J. Qi, X. Zou, J. L. Wang,

T. Sritharan, P. Yang, L. Bellaiche, and L. Chen, Appl. Phys. Lett. **99**, 222904 (2011).

23. F. Johann, A. Morelli and I. Vrejoiu, Appl. Phys. Lett. **99**, 82904 (2011).

24. J. Wang, J. B. Neaton, H. Zheng, V. Nagarajan, S. B. Ogale, B. Liu, D. Viehland, V. Vaithyanathan, D. G. Schlom, U. V. Waghmare, N. A. Spaldin, K. M. Rabe, M. Wuttig, and R. Ramesh, Science **299**, 1719 (2003).

25. Y. H. Chu, Q. Zhan, C.-H. Yang, M. P. Cruz, L. W. Martin, T. Zhao, P. Yu, R. Ramesh, P. T. Joseph, I. N. Lin, W. Tian, and D. G. Schlom, Appl. Phys. Lett. **92**, 102909 (2008).

26. M. F. Wong and K. Zeng, J. Am. Ceram. Soc. **94**, 1079 (2011).

27. S. Yu, D. Sun, W. Yang, and J. Cheng, Mater. Sci. Eng., B **177**, 140 (2012).

28. R. Guo, L. You, M. Motapothula, Z. Zhang, M. B. H. Breese, L. Chen, D. Wu, and J. L. Wang, AIP Adv **2**, 42104 (2012).

29. W. B. Wu, Y. Wang, G. K. H. Pang, K. H. Wong, and C. L. Choy, Appl. Phys. Lett. **85**, 1583 (2004).

30. D. H. Wang, W. C. Goh, M. Ning and C. K. Ong, Appl. Phys. Lett. **88**, 212907 (2006).

31. M. Kumar and K. L. Yadav, J. Appl. Phys. **100**, 074111 (2006).

32. X. H. Wei, W. Huang, Z. B. Yang, J. H. Hao, Scripta Mater. **65**, 323 (2011).

33. G. W. Pabst, L. W. Martin, Y-H Chu, and R. Ramesh, Appl. Phys. Lett. **90**, 72902 (2007).

34. J.S. Wu, C.L. Jia, K. Urban, J.H. Hao, and X.X. Xi, J. Mater. Res. **16**, 3443

(2001).

35. J.-Z. Huang, Y. Wang, Y. Lin, M. Li, and C. W. Nan, J. Appl. Phys. **106**, 063911 (2009).

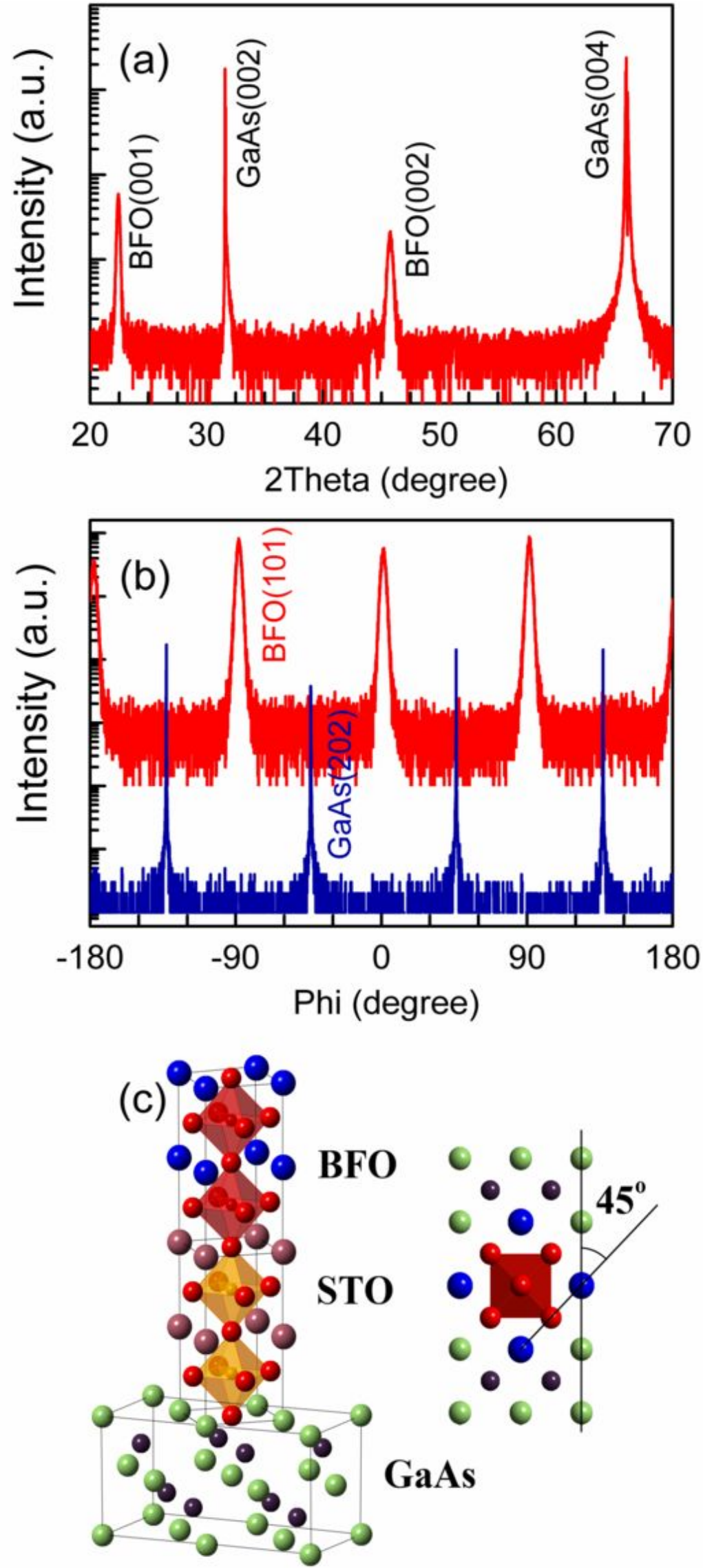

Fig. 1

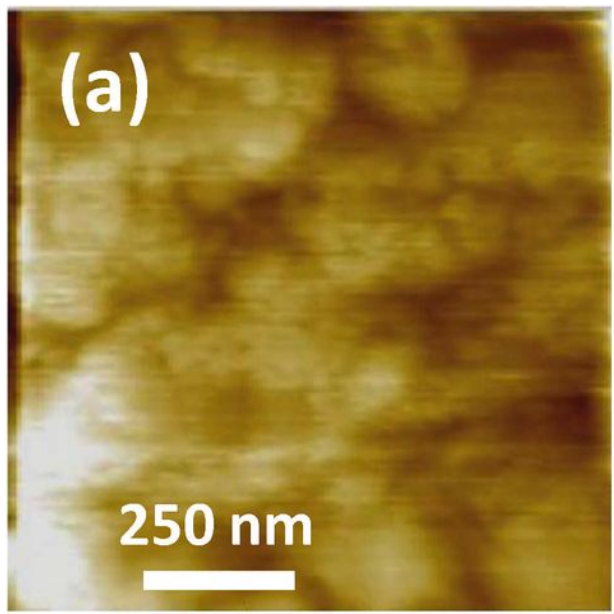
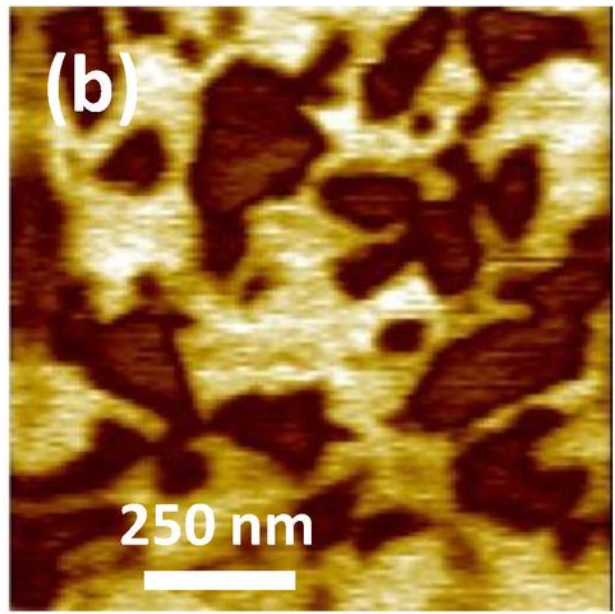
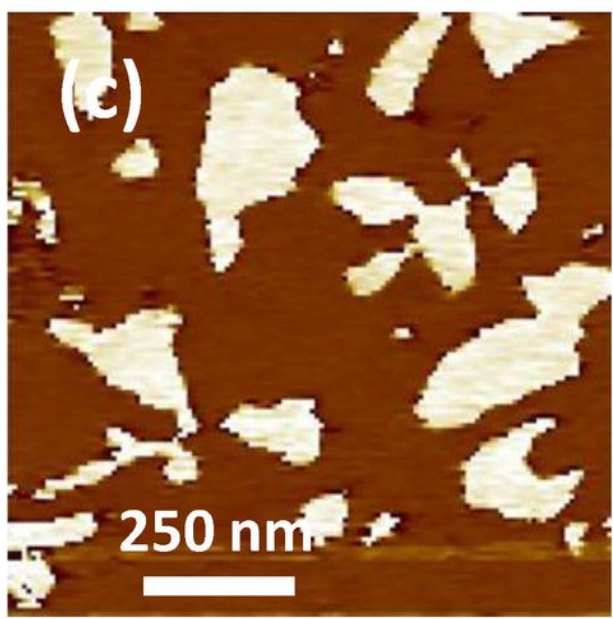
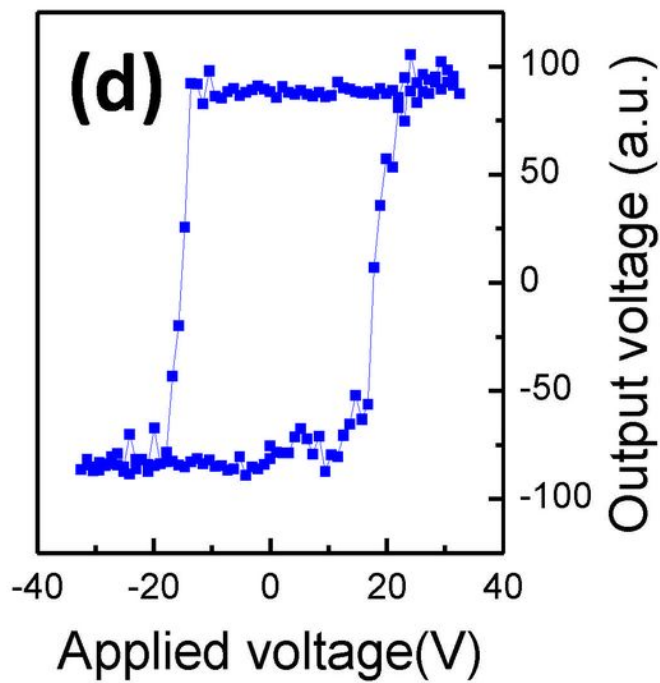

Fig. 2

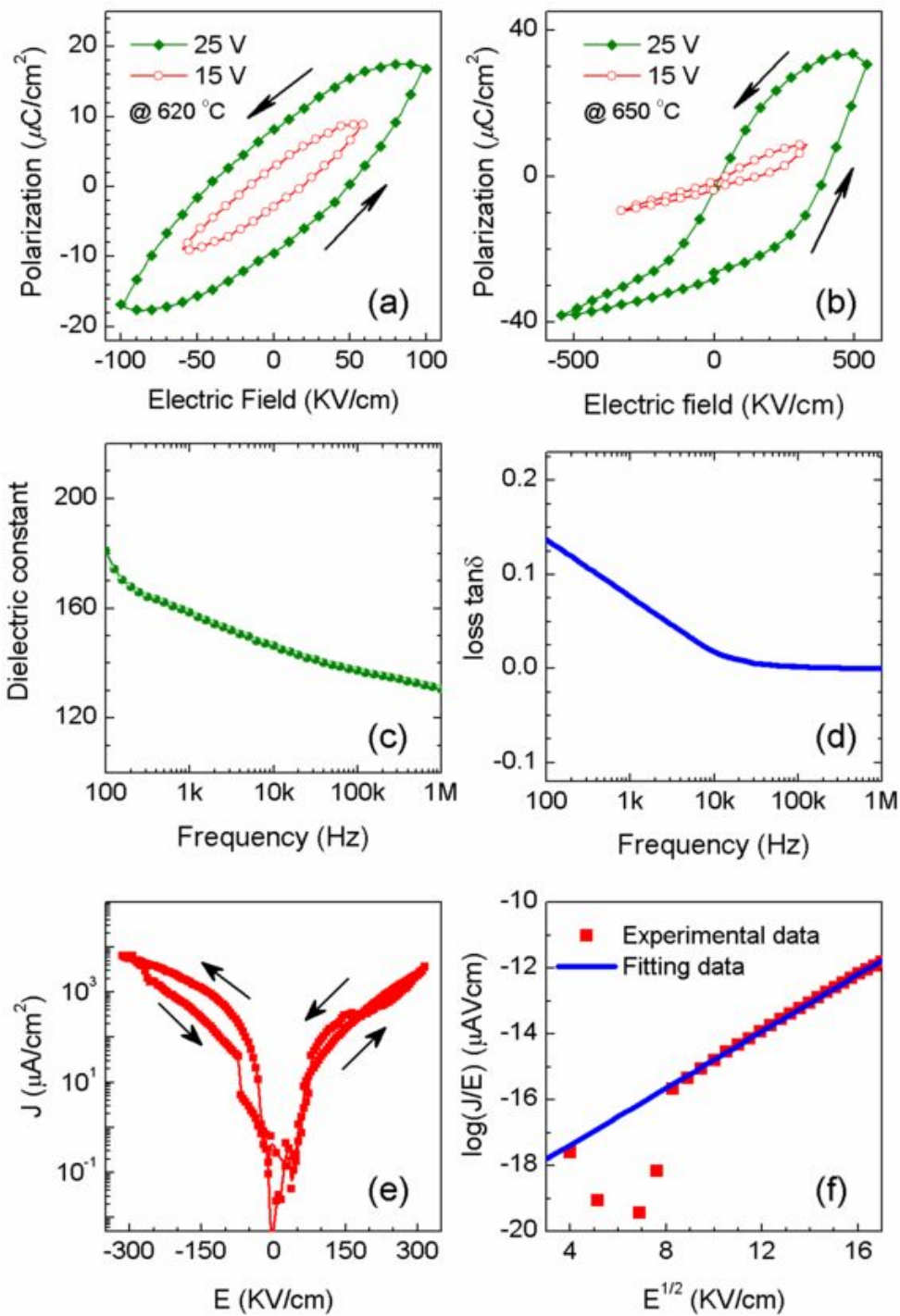

Fig. 3

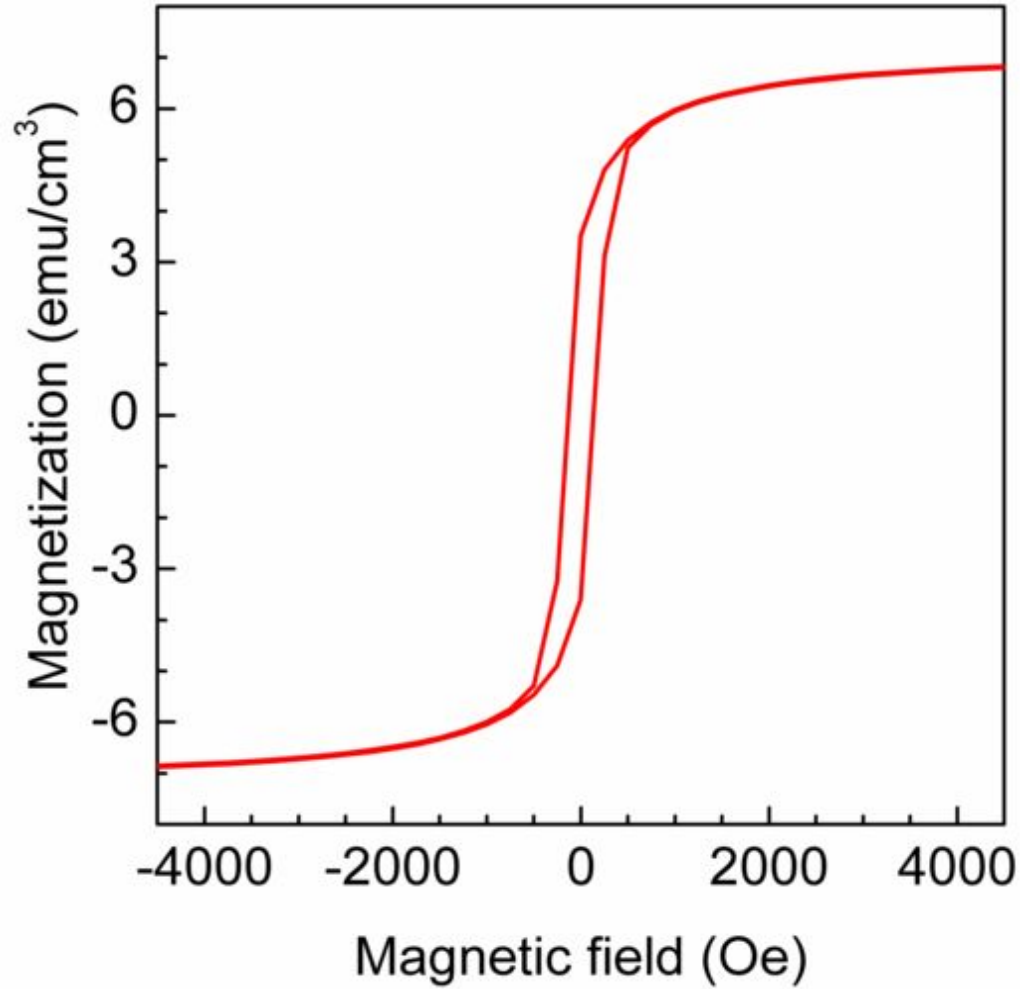

Fig. 4